\documentclass[iop,apj,tighten]{emulateapj}
\usepackage{apjfonts} 
\usepackage{amsmath,amstext}
\usepackage[breaklinks,colorlinks,citecolor=blue,linkcolor=magenta]{hyperref} 
\usepackage[all]{hypcap} 
\usepackage{verbatim}

\begin{document}

\title{{\it Ab Initio} Cosmological Simulations of CR7 as an Active Black Hole}
\author{Joseph Smidt\altaffilmark{1}, Brandon K. Wiggins\altaffilmark{1,2}, Jarrett L. Johnson\altaffilmark{1}}
\affil{$^{1}$Center for Theoretical Astrophysics, Los Alamos National Laboratory}
\affil{$^{2}$Brigham Young University, Department of Physics and Astronomy, ESC, Provo, Utah 84602}

\altaffiltext{*}{jsmidt@lanl.gov}

\begin{abstract}
We present the first \textit{ab initio} cosmological simulations of a CR7-like object which approximately reproduce the observed line widths and strengths. In our model, CR7 is powered by a massive ($3.23 \times 10^7$ $M_\odot$) black hole (BH) the accretion rate of which varies between $\simeq$ 0.25 and $\simeq$ 0.9 times the Eddington rate on timescales as short as 10$^3$ yr. Our model takes into account multi-dimensional effects, X-ray feedback, secondary ionizations and primordial chemistry.  We estimate Ly-$\alpha$ line widths by post-processing simulation output with Monte Carlo radiative transfer and calculate emissivity contributions from radiative recombination and collisional excitation. We find the luminosities in the Lyman-$\alpha$ and He II 1640 angstrom lines to be $5.0\times10^{44}$ and $2.4\times10^{43}$ erg s$^{-1}$, respectively, in agreement with the observed values of $>$ $8.3\times10^{43}$ and $2.0\times10^{43}$ erg s$^{-1}$. We also find that the black hole heats the halo and renders it unable to produce stars as required to keep the halo metal free. These results demonstrate the viability of the BH hypothesis for CR7 in a cosmological context. Assuming the BH mass and accretion rate that we find, we estimate the synchrotron luminosity of CR7 to be $P \simeq 10^{40} - 10^{41}$ erg s$^{-1}$, which is sufficiently luminous to be observed in $\mu$Jy observations and would discriminate this scenario from one where the luminosity is driven by Population III stars. 
\end{abstract}

\maketitle

\section{Introduction}
One of the most intriguing observations of the high-redshift Universe is the detection of the extraordinarily bright Lyman-$\alpha$ emitting object in CR7, which exhibits some of the principal characteristics predicted for a galaxy composed solely of pristine primordial gas (Sobral et al. 2015).  Among these characteristics are bright He II 1640 angstrom emission, indicative of a hot nebula powered by radiation with a particularly hard spectrum (e.g. Bromm et al. 2001; Tumlinson et al. 2001; Oh et al. 2001; Schaerer 2002); the absence of emission features from elements heavier than the helium produced during Big Bang nucleosynthesis; and a nearby ($\ga$ 5 kpc) metal-enriched galaxy, ultraviolet (UV) radiation from which could have suppressed star formation, and the resultant metal enrichment from supernovae, in its progenitor dark matter (DM) halos.   

These characteristics are consistent with two possibilities for the sources of the radiation powering CR7:  a massive ($\ga$ 10$^7$ M$_{\odot}$) cluster of Population III stars formed from the rapid collapse of primordial gas photo-heated by the neighboring galaxy (Johnson et al. 2010; Visbal et al. 2016), and an accreting black hole (BH) with a mass $\ga$ 10$^6$ M$_{\odot}$ (Pallottini et al. 2015) which may drive a strong outflow that imprints features in the Lyman-$\alpha$ line profile (Smith et al. 2016; Dijkstra et al. 2016).  Concerning the latter possibility, it appears likely that the neighboring galaxy may have produced sufficient UV radiation for the BH to have been born with a mass $\sim$ 10$^5$ M$_{\odot}$ via direct collapse (Agarwal et al. 2016; Hartwig et al. 2015); if so, this would confirm the validity of this mode of BH seeding which has been previously invoked to explain the most massive BHs at $z$ $\ga$ 6 (for reviews see Volonteri 2012; Haiman 2013; Johnson \& Haardt 2016; Latif \& Ferrara 2016) as well as some X-ray sources at $z$ $\ga$ 6 (Pacucci et al. 2015).

In this \textit{Letter}, we present the first cosmological hydrodynamic simulations aimed directly at modeling the Lyman-$\alpha$ emitting object in CR7, under the assumption that it is indeed powered by accretion of primordial gas onto a BH which is seeded at higher redshift ($z$ $\sim$ 15).\footnote{While our simulations do not capture every process contributing to the formation of a CR7-like object, starting from the Big Bang, they are {\it ab initio} in the sense that they track the key processes impacting the formation of such an object starting from realistic cosmological initial conditions.}  In the next Section, we describe our {\it Enzo} radiation hydrodynamic simulations of the growth and radiative feedback from the BH.  In Section 3, we present our modeling of the nebular emission powered by the accretion process at the observed redshift $z$ $\sim$ 6.6 of CR7.  Finally, we conclude with a brief discussion of our results in Section 4. 

\section{Problem setup}

The simulations were preformed using {\it Enzo} (Bryan et al. 2013) that contains the necessary modules for hydrodynamics, radiation, primordial chemistry and black hole physics required in this calculation. The initial conditions were generated by MUSIC (Hahn and Abel (2011)) using the Planck 2015 TT,TE,EE+lowP+lensing+ext best fit parameters (Planck collaboration).  MUSIC was run using the Eisenstein \& Hu transfer function with second-order perturbation theory enabled and an initial redshift of $z = 200$.

To find the optimal halo, several 4 Mpc/h boxes with 256$^3$ resolution were generated with MUSIC.  These boxes were then run in Enzo with both baryons and dark matter down to z = 6.6 with one level of AMR refinement everywhere in the box. YT's (Turk et al. 2011) HOP halo finder was used on several redshift snapshots for each box to find a halo with the right final mass and stability properties: since a nested-refinement simulation was to be run, it is important that the halo remain stationary on the grid down to the final redshift with no large mergers from other halos that formed off the nested grid.  One box containing a $3\times10^{10}$ $M_{\odot}$ halo, roughly consistent with estimates from the literature (Agarwal et al. 2016; Hartwig et al. 2015), was chosen to be the simulation described in the rest of this paper.

MUSIC was then rerun with nested grids to create a central fine-grid region, extending 25\% across the box centered on the known halo, with an effective resolution of 1024$^3$ and the same random seeds as before. This central fine-grid region contains $9.11\times 10^{10}$ $h^{-1}$ M$_{\odot}$ of matter with a dark matter particle mass resolution of 4305.74 h$^{-1}$ M$_{\odot}$ and a baryon mass resolution of 803.9 $h^{-1}$ M$_{\odot}$. These initial conditions became the basis for our production run. 

These initial conditions were evolved in {\it Enzo} with 9-species primordial chemistry and cooling (H, H+, He, He+, He++, e-, H2, H2+ and H-) allowing up to nine levels of AMR refinement in the fine-grid region triggered by overdensities in dark matter or baryons and the additional criteria of 32 cells across a Jean's length. Nine levels of refinement represents a spacial resolution of $\sim $ 30 pc/$h$. The simulation was run until our candidate halo reached 10$^{8}$ M$_{\odot}$ around $z \sim 15$. At this point a massive black hole seed was inserted in the center of the halo and radiation feedback using an X-ray spectrum taken from Johnson et. al (2011; see also e.g. Kuhlen \& Madau 2005) was emitted from the seed as feedback. This spectrum was binned into four equal bins representing the first through fourth 25\% bins of energy of that spectrum (With bin centers at 227, 388, 566 and 838 eV). The accretion onto the black hole was regulated by the subgrid alpha disk formalism of DeBuhr et al. (2010) and a Lyman Werner background of J$_{21} = 10^4$ (in units of 10$^{-21}$ erg s$^{-1}$ cm$^{-2}$ Hz$^{-1}$ sr$^{-1}$) was turned on, in order to mimic the radiative feedback from the galaxies nearby CR7 (e.g. Agarwal et al. 2016). 

A few initial black hole seeds were attempted. The reason being that large black hole seeds, the 10$^{4-5}$ M$_{\odot}$ suggested in the direct collapse model, reach 10$^7$ M$_{\odot}$ rather quickly and then loiter in that region with significantly sub-Eddington accretion. In contrast, 10$^3$ black hole seeds take longer to grow to 10$^7$ M$_{\odot}$ and thus are closer to the near-Eddington accretion proposed in the literature.  We find that candidates that are still within a factor of a few from Eddington at $z = 6.6$ have line ratios that match the observations best with our best seed presented in this paper being 3160 $M_{\odot}$.  It must be stressed however that this seed is dependent on the dynamics of the halo chosen.  For example, had we simulated a halo that arrived at 10$^8$ $M_{\odot}$ later on through mergers, a larger black hole seed more like 10$^{4-5}$ M$_{\odot}$ may have been needed.  Given our simplified approach to modeling the Lyman Werner radiation field, which in the CR7 system is likely dominated by nearby star-forming galaxies, our simulations do not capture details of the initial collapse of the primordial gas in the formation of the black hole seed which relate to the anisotropic radiation field produced by nearby sources (e.g. Dijkstra et al. 2008; Ahn et al. 2009; Agarwal et al. 2014; Regan et al. 2014; Visbal et al. 2014; Chon et al. 2016; Habouzit et al. 2016; Valiante et al. 2016) or to the impact of higher energy radiation, which can alter the chemistry of the primordial gas (e.g. Inayoshi \& Omukai 2011; Regan et al. 2015, 2016).  That said, given that a black hole does in fact form, our simulations track the impact of the radiation produced in the accretion process, which we expect to play a dominant role in determining the chemical and dynamical state of the gas in the host halo.

\begin{figure}[t!]\label{fig:radial}
\begin{center}
\begin{tabular}{@{}cc@{}}
\includegraphics[width=\columnwidth]{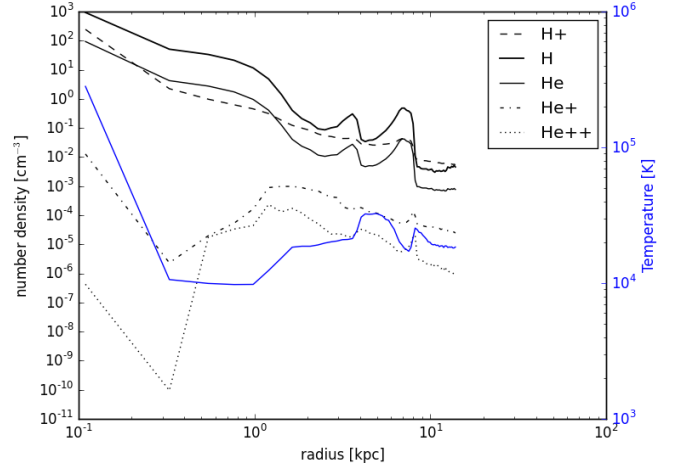}
\end{tabular}
\caption{Azimuthally-averaged number densities for chemical species (black) and gas temperature (blue) as a function of distance from the central BH.}
\end{center}
\end{figure}

Figure~\ref{fig:radial} shows profiles of the halo at $z = 6.6$.

\section{Post-process}

In primordial galaxies, Lyman-$\alpha$ primarily originates from a sequence of de-excitations in atomic hydrogen following recombination. If the ionizing spectrum is harder than that produced by massive Pop III stars, soft x-rays ($\sim 1$ Kev) can escape largely ionized regions and be absorbed in the neutral IGM. Ionizations by high energy photons result in correspondingly energetic free electrons which can serve to also ionize and excite additional neutral hydrogen. Up to 30\% of an electron's energy in such scenarios can result in Lyman-$\alpha$ luminosity (Baek \& Ferrara 2013, Valdes \& Ferrara 2008). We note that this scenario will not take place with a softer ionizing spectrum as ionizing photons are liable to be absorbed in a larger HII region and electron enegies only contribute to heat the gas through electron-electron scatterings. 

\textit{ENZO+MORAY} treats secondary ionizations (Wise \& Abel 2011). The Lyman-$\alpha$ emissivity in the frame of the fluid $\epsilon(\nu)$ of a parcel of gas of volume $V$ with electron and proton number densities $n_e$ and $n_p$ from radiative recombinations is just
\begin{equation}
\epsilon(\nu) = C n_e n_p \alpha_B E_{\mbox{\tiny Ly-$\alpha$}} \phi(\nu) V,
\end{equation}
where $C$ is the fraction of recombinations resulting in the Lyman-$\alpha$ transition, $\alpha_B$ is the case-B recombination coefficient, $E_{\mbox{\tiny Ly-$\alpha$}}$ is the energy of the $2 \rightarrow 1$ transition, and $\phi(\nu)$ is the Voygt line profile which takes into account the effects of thermal doppler broadening and is normalized to 1. The escape fraction $f_{\mbox{\tiny esc}}$ of Lyman-$\alpha$ photons for a BH scenario is $\sim 0.5$ and $\sim 0.1$ for primordial stellar populations respectively (see Hartwig et al. 2015). Our process also takes into account contributions from collisional excitation and the effects of collisional de-excitation. 

\begin{figure*}[t!]\label{fig:pretty}
\begin{center}
\begin{tabular}{@{}cc@{}}
\includegraphics[width=2.0\columnwidth]{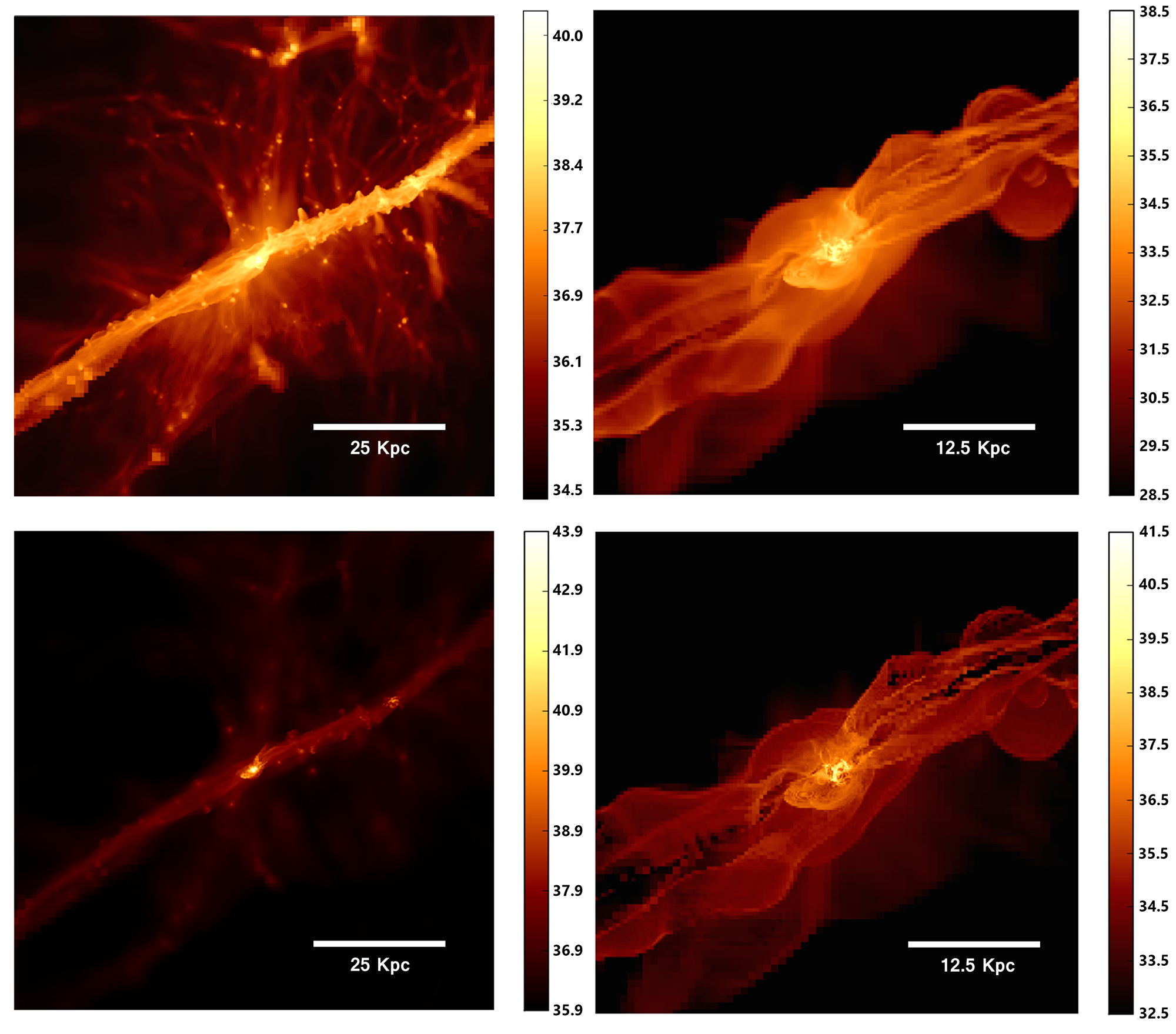}
\end{tabular}
\caption{Projection of logarithm of emissivity [erg s$^{-1}$ kpc$^{-2}$] in Lyman-$\alpha$ (left panels) and HeII 1640 \AA\ (right panels). In the top panels, we plot emissivity due to radiative recombination. In the bottom panels, we plot emissivity due to collisional excitations. Note the difference in color scales.}
\end{center}
\end{figure*}

Lyman-$\alpha$ is a resonate line, so line strengths and widths must be calculated with a method which takes into account scattering. Our Monte Carlo (MC) Lyman-$\alpha$ transfer calculation, facilitated by the HOT framework (Warren \& Salmon 1995) is identical to that found in Djistra (2014) but we do not take into account energy losses from recoil (e.g. Barnes et al. 2014) or relativistic effects which are negligable at these energies. Our MC process further takes into account destruction of Lyman-$\alpha$ via collisional de-excitations and photo-ionization of excited hydrogen which can be effective in metal-free environments (Dijkstra, Gronke \& Sobral 2016). We calculate Lyman-$\alpha$ spectra via a peel-off method (e.g. Zheng \& Miralda-Escud{\'e} 2002) and accelerate the scheme with the prescription in Barnes (2009). We verified our code against the static slab and expanding sphere test cases. We further assume steady-state transfer, i.e. we post-process only a single data dump at $z=6.6$. 

HeII 1640 \AA\ emission may also be recombinatory, but at temperatures $10^{5} - 10^{6}$ K, emission from collisional excitations dominate. To derive HeII 1640 \AA\ emission, we use the method in Yang et al. (2006). Such emission is not resonant, so will only be thermally and kinematically broadened by bulk fluid motions.

\section{Discussion and Conclusions}
\capstartfalse
\begin{deluxetable}{lccccccc}
  \tablecolumns{3}
  \tablecaption{Observations vs. Simulation \label{table:observations}}
  \label{tab_properties}
  \tablehead{     & & \multicolumn{2}{c}{Observations}     && \multicolumn{2}{c}{Simulation} \\  \cline{3-4} \cline{6-7} \\ \colhead{Line} && \colhead{Luminosity} & \colhead{FWHM} && \colhead{Luminosity} & \colhead{FWHM} }
\startdata
Lyman-$\alpha$ && $> 8.3 \times 10^{43}$ & $266 \pm 15$ && $ 5.0 \times 10^{44}$ & 310\\
HeII 1640 \AA\ && $2.0 \times 10^{43}$ & $130 \pm 30$ && $2.4 \times 10^{43}$ & 210
\enddata
\tablecomments{Line luminosities and length widths (FWHM) are given in erg s$^{-1}$ and km s$^{-1}$ respectively. Observed line widths and strengths adopted from Sobral et al. (2015).}
\end{deluxetable}
  \capstarttrue   
    
CR7 is the brightest Ly-$\alpha$ emitter known at $z$ > 6. At a redshift of $z = 6.6$ it is estimated to have a Ly-$\alpha$ luminosity $L_{\mbox{\tiny Ly$\alpha$}} \gtrsim 8.0 \times 10^{43}$ erg s$^{-1}$ with a narrow line width of $\sim 266 \pm 15$ km s$^{-1}$. The HeII 1640 \AA\ line is luminous $L_{\mbox{\tiny HeII}} \simeq 2.0 \times 10^{43}$ erg s$^{-1}$ with a width of $130 \pm 30$ km s$^{-1}$. This implies a large $L_{\mbox{\tiny HeII}}/L_{\mbox{\tiny Ly$\alpha$}} $ ratio $\simeq 0.22$ (Sobral et al. 2015) which is strongly suggestive of a hard ionizing spectrum (Pallottini et al. 2015).

By $z = 6.6$, our halo has acquired $3 \times 10^{10}$ M$_\odot$ and the 3162 M$_{\bullet}$ initial seed has grown to $3.23 \times 10^7$ M$_\odot$.  In our cosmological model, CR7 is situated on a major filament and is fed strongly aspherically (see Figure \ref{fig:pretty}). Recombinatory Lyman-$\alpha$ emissivity originates from the inner $\sim 3$ kpc while HeII 1640 \AA\ emission appears in the direct vicinity of the BH. In Figure \ref{fig:radial}, we plot azimuthally averaged number densities of primordial species as a function of radius and overlay gas temperature in blue. Throughout the halo, temperatures and gas pressures are prohibitive for star formation.\footnote{We note, however, that, using somewhat different prescriptions to model accretion and radiative feedback, Aykutalp et al. (2014) find that star formation may occur in some cases. }

\begin{figure}[t!]\label{fig:lyy}
\begin{center}
\begin{tabular}{@{}cc@{}}
\includegraphics[width=\columnwidth]{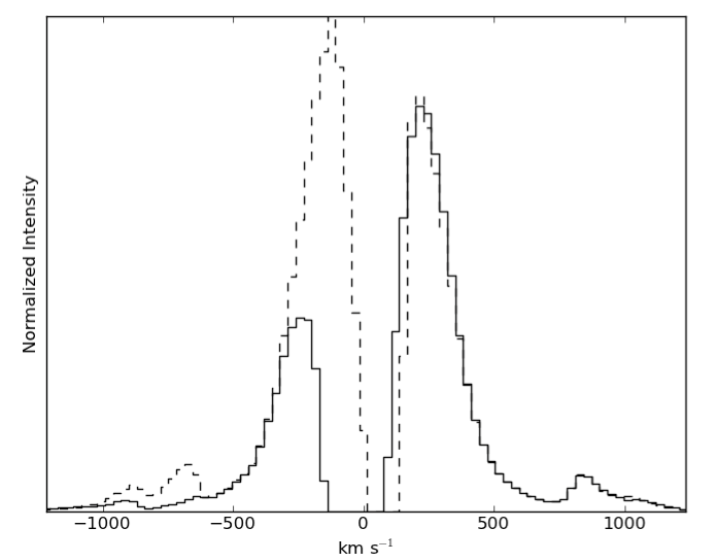}
\label{line}
\end{tabular}
\caption{Lyman-$\alpha$ line profiles for two perpendicular viewing angles of the CR7 model after Monte Carlo radiative transfer post-processing through the simulation domain assuming steady state emission (i.e., we only post-process a single data dump at $z=6.6$), with a peel-off scheme (e.g. Zheng \& Miralda-Escude 2002). The two central, major peaks change size with viewing angle. The spectrum more closely resembling CR7 is emphasized with a solid line.We do not take into account processing through the IGM which may further diminish the blue peak. The larger Lyman-$\alpha$ feature offset redward of the HeII 1640 \AA\ line has a FWHM of $\sim 280$ km s$^{-1}$ in approximate agreement with observations. Though our model predicts a velocity offset, we do not obtain the observed 160 km s$^{-1}$  offset for Lyman-$\alpha$ with respect to the HeII 1640 \AA\ line. (see text).}
\end{center}
\end{figure}

\begin{figure}[t!]\label{fig:bh_accret}
\begin{center}
\begin{tabular}{@{}cc@{}}
\includegraphics[width=0.95\columnwidth]{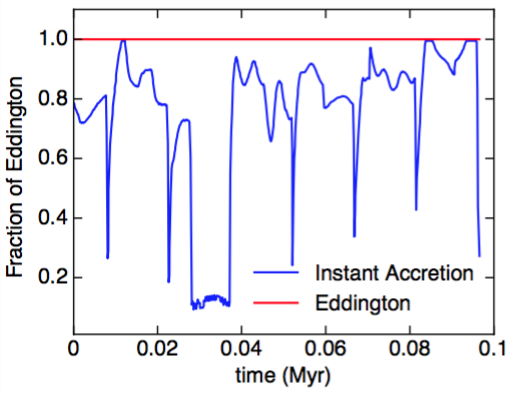}
\end{tabular}
\caption{The instantaneous accretion rate onto the black hole ({\it blue line}) over the 0.1 Myr of the simulation ending at the time of the snapshot we have post-processed at $z$ = 6.6.  In the last 600 yr there is a sharp drop in the accretion rate from roughly the Eddington luminosity down to roughly a fourth of this value.  As the cooling timescale for the Lyman-$\alpha$ emitting gas is of the order of 10$^3$ yr, much of this nebular emission is powered by the accretion luminosity generated at earlier times and is not necessarily relfective of the instantaneous accretion rate.}
\end{center}
\end{figure}

We summarize our results in Table \ref{table:observations}. At at $z = 6.6$, the instantaneous Lyman-$\alpha$ emissivities in our model due to recombination and collisional excitation are $7.6 \times 10^{43}$ and $4.2 \times 10^{44}$ erg s$^{-1}$, respectively.  This would imply a very large fraction $(\sim 60\%)$ of the BH luminosity, $L_{\bullet} = 7.87 \times 10^{44}$ erg s$^{-1}$, being converted into Lyman-$\alpha$. However, the accretion rate in the snapshot we chose to post-process is quite low compared to the near-Eddington accretion rates observed over $10^4$ yr timescales (see Figure \ref{fig:bh_accret}). The accretion rate varies from $\ga 0.9$ to $\simeq 0.25$ Eddington on a timescale of $\simeq 600$ yrs immediately before the post-processed snapshot, which is shorter than the cooling time (of order $10^3$ yrs) for metal-free gas at temperatures and densities of the most Lyman-$\alpha$-luminous zones in the domain. Because of this, the Lyman-$\alpha$ luminosity lags the luminosity of the BH in our calculation. The FWHM of the HeII 1640 \AA\ line is $210$ km s$^{-1}$ which is broader than observations ($130 \pm 30$ km s$^{-1}$) while Lyman-$\alpha$ exhibits a width of $\sim 280$ km s$^{-1}$ after MC post-process which is in rough agreement with observed values of $\sim 266 \pm 15$ km s$^{-1}$ (see Figure \ref{fig:lyy}). This is achieved with $M_{\bullet} = 3.23 \times 10^7$ M$_\odot$, $\dot{M}_{\bullet} = 0.16$ M$_\odot$ yr$^{-1}$ and $L_{\bullet} = 7.87 \times 10^{44}$ erg s$^{-1}$.  We note that the instantaneously low accretion rate of the BH may be reflected in a lower luminosity in continuum photons originating from the disk, which could contribute to raising the observed equivalent width of the HeII 1640 \AA\ line.

The HeII 1640 \AA\ and Lyman-$\alpha$ lines in CR7 are offset in velocity space by $\Delta v = +160$ km s$^{-1}$ (Sobral et al. 2015). 1D calculations by Smith et al. (2016) are able to reproduce this offset by demonstrating that a luminous central source with a hard spectrum drives an outflow which would separate HeII and Ly-$\alpha$ emission in velocity space. Our model does not incorporate feedback from Lyman-$\alpha$, but we are able to reproduce an offset by selecting a fortuitous viewing angle of our 3D model which has a complex bulk-fluid velocity and density field. The relative size of the two Lyman-$\alpha$ peaks varies depending on the orientation of observations. A single Lyman-$\alpha$ feature is observed in CR7, but we do not take into account extinction of the Lyman-$\alpha$ profile through the IGM which could further diminish the blue peak giving the appearance of a single peak displaced redward of the HeII 1640 \AA\ emission (Dijkstra et al. 2016).  The major feature in the spectrum which agrees best with the observed spectrum of CR7 (solid line) is asymmetric and offset with respect to the HeII 1640 \AA\ emission as is observed (Sobral et al. 2015) though our offset (+220 km s$^{-1}$) is larger than CR7's (+160 km s$^{-1}$). An expanding shell of material is not evident in our model. There are several possible explanations for this. Our CR7 model is largely ellipsoidal, not spherical, being compressed along the direction of the angular momentum vector. This strong flattening of the system allows for radiation to escape CR7 through higher angles of latitude.

We note that Population III stars are expected to produce nebular Lyman-$\alpha$ emission principally from recombination, and not from collisional excitation as is the case for the harder X-ray spectrum emitted from an active black hole.  We estimate an upper limit for the Lyman-$\alpha$ luminosity of 1.14 $\times 10^{44}$ that could be generated by recombination emission powered by Population III stars, by assuming both that the gas within the halo is fully ionized.\footnote{For this estimate we have adopted the density field extracted from our simulation.  If the gas is significantly more dense in the halo in the case of Population III star formation, then the luminosity in recombination lines could be higher due to the higher recombination rate.  That said, the strong photoionization feedback from massive Population III stars would likely drive down the density of the gas very quickly after the formation of the stars (e.g. Whalen et al. 2004).}  This is only luminous enough to explain the CR7 luminosity if the Lyman-$\alpha$ escape fraction is $f_{\mbox{\tiny esc}} \ga 0.8$, much higher than previously estimated (e.g. Hartwig et al. 2015).  This, along with the extremely high star formation efficiency of $\ga$ 0.1 that is required for Population III stars to explain the observed emission (Visbal et al. 2016; Smith et al. 2016), poses a strong challenge to this alternative model for CR7.

While Ly-$\alpha$ is extended on the kpc scale, much of the helium emission in our model originates from the vicinity of the BH. We find that to maintain line widths $\sim 100$ km/s in agreement with CR7 requires that the line be only broadened by virial velocities of bulk fluid motion and thermal broadening. Our models with more massive BHs both diminish and broaden HeII 1640 \AA\ emission, in some cases giving rise to a double-peaked profile which characterizes line observations of expanding shells of gas. The complex line profiles are not observed in CR7 which suggests an upper limit on BH luminosity in models of CR7 powered by an accreting BH.

Could CR7 be detected in the radio? Our estimate of the accretion rate suggests the BH is accreting within the ``thin disk'' regime or ``quasar mode'' (e.g. Dubois et al. 2011) where the prospects of launching a relativistic jet and powering radio lobes remains unclear. Yet radio synchrotron emission could originate on scales of 100s of Schwarzschild radii from relativistic accretion shocks. For our values of $M_{\bullet}$ and $\dot{M}_{\bullet}$ in the analytic model by Ishibashi \& Courvoisier (2011) and adopting the authors' estimates for electron spectral index $p$ and ambient magnetic field $B$, we estimate a synchrotron luminosity of $\sim 1.0 \times 10^{39}$ erg s$^{-1}$ arising from subgrid scales. The total observed radio flux density $S_\nu$ of CR7 integrated over the source is 

\begin{equation}
S_{\nu} = \frac{P_{\nu_0}}{\nu_0 \cdot 4 \pi D_L^2 (1+z)^{\alpha -1}}
\end{equation}
where $P_{\nu_0}$ is the power radiated at frequency $\nu_0$, $D_L$ is the luminosity distance, $\nu$ is the frequency of observations related to the emitted frequency $\nu_0$ by the doppler shift. We take $\alpha \approx 1.5$, accounting for the $z \sim \alpha$ correlation for steepening radio spectral indices with increasing redshift ($\alpha \simeq 0.7$ at $z = 0$; see Cavagnolo et al. 2010; Condon 1992) for emitted frequencies at 10.0 GHz. For sensitivites of a few $\mu$Jy per beam in L-band ($\nu \simeq 1.3$ GHz), this imposes a detection limit of $P_{\nu_0} \gtrsim 10^{41}$ erg s$^{-1}$. This quantity is mildly sensitive to the choice of $\alpha$; for local-universe values, the detection limit becomes $P_{\nu_0} \gtrsim 10^{40}$ erg s$^{-1}$.  

Scaling relationships for accretion, jet power and radio luminosity exist in the literature. If we insert our active galactic nuceli (AGN) luminosity in X-rays and M$_{\bullet}$ into a model by Merloni et al. (2003) which is based on a thoery of scale-invariant jets (Heinz \& Sunyaev 2008), we obtain a log-mean radio luminosity of $P \approx 1.4 \times 10^{40}$ erg s$^{-1}$ though their model suffers from very large scatter, spanning a few orders of magnitude.  If we apply our data to the models of Meier (2001), adopting the thin disk regime for the Schwarzschild case (his equation 4), we obtain a jet power of $\sim 1.0 \times 10^{41}$ erg s$^{-1}$. If we utilize the relation between jet power and radio output found in Cavagnolo et al. (2010), this corresponds to a radio luminosity of only $10^{35}$ erg s$^{-1}$.
Our BH is accreting at an appreciable fraction of Eddington, and such disks can be unstable and oscillate between low, hard (LHS) state where radio brightness is expected to increase and high, soft (HSS) state. At a reduced accretion rate of 0.1 Eddington, jet power for a rotating black hole would be $\gtrsim 10^{44}$ erg s$^{-1}$ corresponding to a radio luminosity of $P_{\nu_0} \simeq 2.0 \times 10^{42}$ erg s$^{-1}$ which would be detectable in $\mu$Jy observations. Such jets may drive outflows and could be responsible for the 160 km s$^{-1}$ offset of the Ly-$\alpha$ line from the systemic velocity. In our model, the BH accretes at a large fraction of Eddington for a large duration of the simulation, which may correspond to a thicker, radiation-dominated disk, greater probabilities of launching a jet, and hence larger radio luminosities. Any luminous radio emission from CR7 is likely to be evidence of AGN activity, as synchrotron emission in star formation regions primarily arises from supernova remnants which are thought to be absent in CR7 from the lack of observed metals.

Our work provides important additional support for the massive BH model of CR7 via detailed radiation hydrodynamic cosmological simulations.

\acknowledgments{
  We are grateful to Ryan Wollaeger, Michael Murillo and Joshua Dolence for useful discussions which improved the quality of this work. We are grateful for the insight and comments provided by the anonymous reviewer.

\bibliographystyle{yahapj}

\end{document}